\documentclass[manuscript]{aastex631}

\begin{document}

\title{The ultraviolet \ion{C}{2} lines as a diagnostic of  $\kappa$-distributed electrons
   in planetary nebulae}

\correspondingauthor{Yong Zhang}
\email{zhangyong5@mail.sysu.edu.cn}

\author{Zheng-Wei Yao}
\affiliation{School of Physics and Astronomy, Sun Yat-sen University, Zhuhai, 519082, People’s Republic of China}

\author[0000-0002-1086-7922]{Yong Zhang}
\affiliation{School of Physics and Astronomy, Sun Yat-sen University, Zhuhai, 519082, People’s Republic of China}
\affiliation{CSST Science Center for the Guangdong-Hongkong-Macau Greater Bay Area, Sun Yat-Sen University, Guangdong Province, People’s Republic of China}
\affiliation{Laboratory for Space Research, The University of Hong Kong, Hong Kong, People’s Republic of China}

\begin{abstract}

Non-Maxwellian $\kappa$ electron energy distributions (EEDs) have been proposed in recent years to resolve the so-called ``electron temperature and abundance discrepancy problem'' in the study of planetary nebulae (PNe). Thus the need to
 develop diagnostic tools to determine from observations
 the EED of PNe is raised.  Arising from high energy levels, 
  the ultraviolet (UV) emission lines from PNe present intensities that depend sensitively  on the high-energy tail of the EED.  In this work, we investigate the feasibility of 
using the \ion{C}{2}]$\lambda$2326/\ion{C}{2}$\lambda$1335 intensity ratios as a diagnostic of the deviation of the EED from the Maxwellian distribution (as represented by the $\kappa$ index). We use a Maxwellian decomposition approach to  derive the theoretical $\kappa$-EED-based collisionally excited coefficients of \ion{C}{2}, and then compute the \ion{C}{2} UV intensity ratio as a function of the $\kappa$ index. We analyze the archival spectra acquired by the {\it International Ultraviolet Explorer} and measure the intensities of \ion{C}{2} UV lines  from 12 PNe. By comparing the observed line ratios and the theoretical predictions, we  can infer their $\kappa$ values. With the Maxwellian-EED hypothesis, the observed \ion{C}{2}]$\lambda$2326/\ion{C}{2}$\lambda$1335 ratios are found to be generally lower than those predicted from the observed optical spectra. This discrepancy  can be  explained in terms of  the $\kappa$ EED. Our results show that the $\kappa$ values inferred range from 15 to infinity, suggesting a mild or modest deviation from the Maxwellian distribution. However, the $\kappa$-distributed electrons are unlikely to exist throughout the whole nebulae. A toy model shows that  if just about 1--5 percent of the free electrons in a PN had a $\kappa$-EED as small as  $\kappa=3$, it would be sufficient to account for the observations.

\end{abstract}



\section{Introduction}
Photoionized gaseous nebulae, including planetary nebulae (PNe) and \ion{H}{2} regions, serve as a vital probe of the chemical evolution of galaxies and as a natural laboratory for understanding the plasma physics found in extreme physical
environments. Extensive observational studies of nebulae
have revealed stark differences between  the electron temperatures and elemental abundances 
derived from   collisionally excited lines (hereafter CELs) and those derived from
recombination spectra, which is often called the
``electron temperature and abundance discrepancy problem''
\citep[e.g.,][]{Peimbert, Liua, Zhanga,gar17}. Several alternative solutions have been previously proposed
and continue to be hotly debated \citep[see e.g.,][as examples of the recent contributions to this debate]{Danehkar,Hajduk,G-R,Ruiz-Escobedo}.
 Two mechanisms often cited are
temperature fluctuations (characterized by the $t^2$ temperature fluctuation parameter) and the two-component model consisting of hydrogen-deficient clumps distributed within a diffuse nebula.

Recently, a new promising scenario has been presented for solving the problem. 
The determinations of nebular temperatures and abundances in most published literature are based on a fundamental but
insufficiently demonstrated assumption that the electron energy distribution (hereafter EED) necessarily follows a 
Maxwell-Boltzmann (hereafter MB) energy distribution. If the EED was indeed non-thermal, the inconsistencies of current plasma diagnostics would be essentially resolved
 \citep{Nichollsa, Nichollsb, Doptiab}, in which case the EED
becomes characterized by a $\kappa$ function where a
lower $\kappa$ index corresponds to a larger departure from the MB distribution.
Despite the plausibility of this solution, the mechanism
for energizing non-thermal electrons is so far unknown. Much of the criticism 
to the $\kappa$-EED hypothesis is from a theoretical
perspective. Given the relatively short relaxation timescale of non-thermal electrons,
the photoionization process of gaseous nebulae is expected not
 to generate a $\kappa$ EED
\citep{Ferland,Draine}, casting serious doubt on this hypothesis. Nevertheless, it is premature to
conclude that the currently proposed alternative
 scenarios are necessarily valid. For instance, if
  the hypothesis of temperature fluctuations held, the $t^2$ values
required to explain the observations would be too large to be reproduced by
canonical photoionization models.
Regarding the two-component model, the proposed but unconfirmed 
 hydrogen-deficient inclusions are also of
 unknown origin.

The $\kappa$ distribution was first introduced to provide an empirical fit to the energy distribution of
the earth's magnetospheric plasma \citep{Vasyliunas}, and was then found to be a natural consequence 
of the Tsallis statistical mechanics \citep{Livadiotisa}.
It has been widely found to be present in the
solar wind \citep{Pierrard}, in the heliosheath \citep{Livadiotis}, and the solar corona \citep{Nicolaou}.
The proposed origin for particle acceleration, such as magnetic reconnection, is a hot topic in space plasma research. 
It is not unreasonable to speculate that such processes may also affect the EED of PNe, where
the ``electron temperature and abundance discrepancy problem'' is more prominent than observed in \ion{H}{2} regions.

Spectral observations may allow us to confirm or refute the presence of the $\kappa$ EED in PNe. The \ion{C}{2} dielectronic recombination lines
and the continuum emission spectra near the Balmer jump (hereafter BJ)
have been proposed to determine the EED
\citep{ss13,ss14,Zhangb}. However, primarily due to large uncertainties, these results  are inconclusive or even conflicting. 
\citet{Zhangc} determined the $\kappa$ index for a sample of PNe by comparing the CEL ratios and BJ intensities, which 
appears to not correlate with the other properties of PNe.
\citet{Lin} found that the introduction of the $\kappa$ EED can bring
in line 
the intensity ratios of the \ion{O}{2} recombination lines
and those predicted by [\ion{O}{3}] CEL observations, 
but the inferred  $\kappa$ values are not in agreement with
those obtained from  comparison of the CEL and BJ intensities.
Such studies based on optical spectroscopy suggest that the non-thermal electrons are unlikely to be uniformly 
distributed across the whole nebulae,
but they cannot exclude their presence on a small spatial scale as is found
to be the case for the solar system plasma.

The non-thermal high-energy tail of the $\kappa$ EED would cause a
stronger impact on the excitation of the ultraviolet (UV) emission
lines than on the optical lines. Therefore,  UV line fluxes 
can in principle serve
as an excellent probe of the EED in PNe.
Based on the observed \ion{C}{2} UV
line ratios, \citet{Humphrey} and \citet{Morais} found evidences supporting the existence of the $\kappa$ EED in  
Type 2 Active Galactic Nuclei (AGNs), the high-energy analogue of PNe.

In this paper, we investigate the impact of the $\kappa$ EED on
the  \ion{C}{2}]$\lambda$2326/\ion{C}{2}$\lambda$1335 line ratios,
aiming at examining whether the UV observations indicate the
presence of $\kappa$-EED in PNe. Section~2 describes the archival data we
will be using. Section~3 describes the method used to derive
the \ion{C}{2} UV line ratio under a $\kappa$ EED.
The results are presented in Section~4 and are discussed in Section~5. Section~6 summarizes the conclusions.



\section{Sample and observational data}
 
 The UV observations of PNe have been obtained from the {\it International Ultraviolet Explorer} (IUE).
 We extracted the low-resolution  spectra obtained with the Short Wavelength Prime (SWP) and Long Wavelength Prime/Long Wavelength Redundant (LWP/LWR) cameras from  
 the INES (IUE Newly Extracted Spectra) Archive\footnote{http://sdc.cab.inta-csic.es/cgi-ines/IUEdbsMY}.
 The SWP and LWP/LWR spectra cover the following wavelength ranges:  1150--1980 and 1850--3350\,{\AA}, which therefore include the
 \ion{C}{2}$\lambda$1335 and  \ion{C}{2}]$\lambda$2326 lines, respectively.
For the sample selection, we focused on the PNe studied by \citet{Zhangc} for the sake of comparison. 
Due to the severe extinction at UV wavelengths, The \ion{C}{2} UV lines for most of the PNe are faint or invisible. 
 Discarding those nebulae with only one of the 
 \ion{C}{2} lines being detected, we selected the PNe  which possess a
 reasonably good signal-to-noise ratios.
In order to ascertain whether the observed features are indeed \ion{C}{2} lines rather than glitches or
other contaminations,
 we carefully compared the widths and peak velocities of the two
 features in each PN, and excluded those with apparent inconsistencies.
  Finally, we ended with the 12 PNe listed in Table~\ref{tab1}.
  These PNe are ionized bounded and show relatively high electron temperatures, which are the physical conditions favoring the emission
 of \ion{C}{2} UV CELs.
 All IUE observations were taken with a large aperture of $10\arcsec\times23\arcsec$ except for two PNe.
For NGC\,6572, the LWR spectrum obtained with the small aperture  ($3\arcsec$ diameter) was used.
We did not attempt to make an aperture correction since NGC\,6572 is a compact young PN.
 For  NGC\,3918, both spectra  were obtained using the smaller aperture.

 We subtracted the continuum emission from the IUE spectra by performing a linear fit to the line-free spectral regions. 
 The fluxes of the two \ion{C}{2} lines were measured using the Gaussian line profile fitting. Figure~\ref{Fig.1} shows the IUE spectra and the line fittings for
 the 12 PNe. The integrated fluxes $F(\lambda)$ were then deredened 
 to derive the intensities by $I(\lambda)$ = 10$^{c({\rm H}\beta)f(\lambda)}$ $F(\lambda$), where $c({\rm H}\beta)$
 is the logarithmic extinction constant taken from \citet{Cahn},
 and $f(\lambda)$ is the standard Galactic extinction law with $R_V = 3.1$ \citep{Howarth}. See Table~\ref{tab1} for the details.


 
 \section{Methodology} \label{meth}
 
 \citet{Nichollsa} for the first time introduced 
the $\kappa$ function to describe the EED of PNe, which
consists of a cold
 MB-like core characterized by a MB temperature of
 $T_c$ and a power-law superthermal tail,
\begin{equation} 
     f_{\kappa}(E, T_{U}, \kappa)= \frac{2}{\sqrt{\pi}}\frac{\Gamma(\kappa+1)\sqrt{E}}{(k_BT_{U})^{1.5}(\kappa-1.5)^{1.5}\Gamma(\kappa-0.5)}
     \left[1+\frac{E}{(\kappa-1.5)k_BT_{U}}\right]^{-\kappa-1} \,,
   \end{equation}
where $E$, $k_B$ and $\Gamma$ are the electron energy, Boltzmann constant, and the Gamma function, respectively.
The dimensionless $\kappa$ index ranges from $1.5$ to $\infty$.
 $T_{U}$ is the nonequilibrium temperature representing the mean kinetic energy of free electrons,
 which satisfies the relation 
 \begin{equation} \label{tc}
 T_U= \frac{\kappa}{\kappa-1.5} T_c \,. 
 \end{equation} 
When $\kappa\to\infty$, the formula decays to a MB form with $T_U=T_c$.

The theoretical intensity ratios of \ion{C}{2} CELs can be computed by solving the ten energy level balance equation.
For that purpose, the collisionally excited rate coefficients ($q$)
between any two energy levels  are  derived from
 the energy-dependent collision cross sections, $\sigma(E)$, integrated over the EED function. This means
 \begin{equation} \label{rate}
     q_M(T_M)=\int \sigma(E) \sqrt{E}f_M(E, T_M)dE
   \end{equation}
under the MB-EED hypothesis
(hereafter, we use the subscripts `$M$' and `$\kappa$' to distinguish between
the cases under the MB or the $\kappa$ EED assumption, respectively).
The tabulated $q_M(T_M)$ values are available in \citet{Blum}.
In principle, $q_{\kappa}(T_{U}, \kappa)$ could be obtained by substituting $f_{\kappa}(E, T_{U}, \kappa)$ for $f_M(E, T)$ in Equ.~\ref{rate}, but such
procedure is impractical because numerical evaluation of $\sigma(E)$ is not available from the literature.
 \citet{Hahn} proposed an easier approach to determine  $q_{\kappa}(T_{U}, \kappa)$ 
 using the tabulated  $q_M(T_M)$.
 The $\kappa$ function can be mathematically decomposed into several weighted MB functions,
\begin{equation} \label{fkappa}
      f_{\kappa}(E, T_{U}, \kappa)= \sum_{j}c_j (\kappa)f_M(E, T_{M,j}) 
   \end{equation}
with  $T_{M,j}=a_j T_{U}$, where
$a_j$ and  the weighting factor $c_j$ are independent of $T_U$.
 Equs.~\ref{rate} and \ref{fkappa} indicate that
 \begin{equation} 
      q_{\kappa}(T_U, \kappa)= \sum_{j}c_j (\kappa)q_M(T_{M,j})  \,.
   \end{equation}
Therefore, we can obtain $q_{\kappa}(T_{U}, \kappa)$ by using the above formula and
the $a_j$, $c_j$, and  $q_M(T_M)$ values provided by
\citet{Hahn} and   \citet{Blum}. This approach for instance has been successfully 
employed to develop diagnostics for the solar plasma
\citep{dz21}.
 \cite{dz21} noted that this approach provides highly precise results
with an uncertainty comparable to the typical uncertainty inherent in the
atomic calculations.

The lowest seven levels of C$^{+}$  relevant to the 
\ion{C}{2} UV lines are sketched in Figure~\ref{Fig.2}. As shown in this figure,
 the $\lambda1335$ and $\lambda2326$ features are respectively composed of three and five blended lines
 that cannot be resolved with IUE spectra. In this work, we will focus on the intensity ratios between
 the $^{4}$P$_{5/2}$--$^{2}$P$_{3/2}$  transition at 2326.12\,{\AA} and the $^{2}$D$_{3/2}$--$^{2}$P$_{1/2}$
transition at 1334.52\,{\AA}, which is hereafter called
 $r$(\ion{C}{2}), in short.
Figure \ref{Fig.3} shows $r$(\ion{C}{2}) as a function
of both $T_{U}$ and electron density ($N_e$) at various $\kappa$ 
 values. It is conceivable that
 $r$(\ion{C}{2}) decreases with increasing $T_{U}$ and decreasing
 $\kappa$, which means that higher-energy electrons enhance
 the $\lambda1334.52$ line to a larger degree than
 the $\lambda2326.12$ line.
 An inspection of this figure also suggests that 
 $r$(\ion{C}{2}) is essentially independent of $N_e$. 
 This is because  the two lines have critical densities much higher
 than the typical density found in PNe. At a temperature of $10^4$\,K,
 the $\lambda2326.12$ and $\lambda1334.52$ lines show a critical density
 of $7.3\times10^{8}$\,cm$^{-3}$ and $2.2\times10^{15}$\,cm$^{-3}$,
 respectively. Consequently, once $T_U$ is known, we can infer
 the $\kappa$ index from the observed $r$(\ion{C}{2}).
 At lower $T_U$, $r$(\ion{C}{2}) is more sensitive to
 $\kappa$.
 

The observed $r$(\ion{C}{2}) can be derived by considering the
theoretical relative intensities between the blended lines.
The $\lambda1334.52$ and $\lambda1335.66$ lines arise from the same upper level, and thus their
intensity ratio is strictly equal to 5.1, while $I(\lambda1335.71)$ is about
$1.8\times I(\lambda1334.52)$.  It follows that
the $\lambda1334.52$ line has an intensity of 
0.34 times that of the blended feature.
The $\lambda2326.12$ line is the strongest among the five blended lines, with an intensity of 0.50 times that of 
the whole blended feature.
These relative intensity ratios are extremely insensitive to the nebular physical conditions in the sense that
the blended lines arise from the upper levels with very similar excitation energies and 
high critical densities ($>10^7$\,cm$^{-3}$).
After deblending the \ion{C}{2} $\lambda1335$ and  $\lambda2326$ features, the observed $r$(\ion{C}{2}) are  derived and their values are
 listed in the  7th column of Table~\ref{tab1}.

 \section{Results}


The theoretical and observed $r$(\ion{C}{2}) values are compared in 
Figure~\ref{Fig.4}, where we simply assume $N_e=10^3$\,cm$^{-3}$.
For that we have derived the $T_U$ based on the BJ temperatures, $T_M$(BJ), given in the literature (see the 9th column of Table~\ref{tab1}).
Because the BJ mostly `sees' the cold MB core of the $\kappa$ EED, 
it is a good approximate to assume $T_c=T_M({\rm BJ})$. Then
$\kappa$ and $T_U$ can be iteratively determined from $r$(\ion{C}{2}) and Equ.~\ref{tc}. Their values are presented in the
11th and 13th columns of
Table~\ref{tab1}. The uncertainties of $\kappa$ are found to increase with increasing $T_U$ and $\kappa$.

If the EED obeyed the MB distribution, the data points would overlay
the $\kappa = \infty$ curve in Figure~\ref{Fig.4}.
On the contrary, if the free elections got closer
to the $\kappa$ EED with a smaller $\kappa$ index,
the CEL with a higher excitation energy would be relatively enhanced, and thus
$r$(\ion{C}{2}) would decrease and $T_M({\rm BJ})$ would be lower
than $T_U$, meaning that the data points would shift towards
the left and down in this figure. We found this to be
the case for most of the PNe of our sample, thus providing us a favorable evidence for the $\kappa$ EED in PNe. 
We could derive well constrained $\kappa$ values for six PNe. 
NGC\,6572 shows a low $\kappa$ index of 17. However, the small-aperture of its LWR spectrum may have caused an underestimation of the
\ion{C}{2}]$\lambda2326$ line flux, and thus likely resulted in an  underestimation 
of the $\kappa$ index.
 The medium $\kappa$ value for the PN sample turns out to be larger than 50, suggesting that most PNe  exhibit only mild to modest deviations 
from the MB distribution.

\citet{Humphrey} found that the $r$(\ion{C}{2})  ratio in their
AGN sample suggests the presence of 
a $\kappa$ EED. We reexamined their results in 
Figure~\ref{Fig.4} using the observational data retrieved from \citet{Vernet}, where we have assumed
 a typical electron temperature of
$1.5\times10^{4}$\,K \citep{Osterbrock}.
Most AGNs exhibit a prominent $\kappa$ EED unless they
have an abnormally high temperature,
in agreement with the conclusion of \citet{Humphrey}.
Figure~\ref{Fig.4} shows that the medium $\kappa$ value 
inferred from the
AGN spectra is around 20.
Recently, through  photoionization model fitting
to a large sample of type 2 AGNs,
\citet{Morais} found that, for most of the objects, 
a $\kappa$ EED with  $\kappa=5$
 provides a better fit to the observed spectra
than the MB EEDs. Such a low $\kappa$ value
suggests a generally larger departure 
from the MB distribution in AGNs than in PNe. 
This could probably be attributed to their high-energy physical condition.
\citet{Humphrey} also indicated that shocks can decrease $r$(\ion{C}{2}) to
$<5$ and therefore account for the measured values of their AGN sample except for one object.
As can be inferred from Figure~\ref{Fig.4}, however,  shock models
cannot reproduce the PN observations.

An effect we did not consider is the continuum fluorescence through
the \ion{C}{2}$\lambda$1335 resonance line, which may either enhance or suppress
the $r$(\ion{C}{2}) ratio, depending on the nebular geometry
\citep[see  discussion in][]{Humphrey}. However, almost all our
PNe show a measured $r$(\ion{C}{2}) ratio lower relative to the MB values
(Figure~\ref{Fig.4}). Moreover,  the stellar UV continuum luminosity of the PNe 
is presumably low within  the C$^+$ regions due to geometrical dilution. This would
hinder continuum fluorescence as being the cause of the $r$(\ion{C}{2}) discrepancy.

In order to facilitate the evaluation of plasma diagnostics using $\kappa$-EED
calculations appropriate to photoionized gaseous nebulae, 
we present in Table~\ref{A1}
the theoretical $r$(\ion{C}{2}) calculated for various $\kappa$ and $T_U$  values.
Although we  assumed $N_e=1000$\,cm$^{-3}$ in the calculations, the results 
are essentially insensitive to the electron density.
Once $T_c$ using Equ.~\ref{tc} is derived from the BJ,
the $\kappa$ and $T_U$ values can be derived from the observed  $r$(\ion{C}{2}) ratio through  iteration. 
Alternatively, one could use the CELs instead of the BJ for the determination of 
$\kappa$ and $T_U$,  in which case the relations between $T_U$ and $T_M$(CELs) proposed by \citet{Nichollsb} can be used.

 \section{Discussion }
 
We have investigated the feasibility of invoking
the $\kappa$ EED to solve the ``electron temperature
and abundance discrepancy problem'' encountered in PN studies. 
The electron temperature of the singly ionized {C}$^+$ regions can be evaluated 
using the temperature inferred from the 
  [\ion{N}{2}] lines, $T_{M}$(\ion{N}{2}). We have retrieved $T_{M}$(\ion{N}{2}) from
the literature (Table~\ref{tab1}), which is generally found to be higher
than $T_M({\rm BJ})$. 
Within the scenario of  $\kappa$ EED,   $T_{M}$(\ion{N}{2})   strongly depends on the electronic high-energy tail, while 
the BJ depends mostly on the cold core of the distribution.
It is interesting to note
that the derived $T_U$ values consistently lie between $T_{M}$(\ion{N}{2}) and $T_M({\rm BJ})$, suggesting that
the temperature discrepancy  can indeed be alleviated under the
$\kappa$ EED  hypothesis.
The ratio between the O$^{2+}$/H$^+$ abundance derived 
from recombination lines and that derived from CELs, the
so-called  abundance discrepancy factor (ADF),
is introduced to quantify the degree of  abundance discrepancy.
As listed in Table~\ref{tab1}, the ADFs of our sample PNe
range from 1.2 to 5.4. It is unfortunate that those nebulae with extremely large ADFs do not have \ion{C}{2} UV lines detected.
In Figure~\ref{Fig.5}, we show that the $\kappa$ index negatively
correlates with both $T_{M}$(\ion{N}{2})$-T_{M}$(BJ)
and ADFs.
This is compatible with the expectation that a larger departure
from the MB EED should lead to larger temperature and
abundance discrepancies.
However, the correlation coefficients are only moderate, suggesting
that the situation is not as  simple as described here.

If the PN is homogeneous, that is, is presenting a constant   EED, the $\kappa$ indices deduced from different diagnostic line ratios should
have an identical value. 
For the purpose of comparison, Table~\ref{tab1} lists the $\kappa$ indices obtained by \citet{Zhangc} using the [\ion{O}{3}]$\lambda\lambda$4959,5007/$\lambda$4363 ratio and the \ion{H}{1} BJ. The agreement is relatively marginal. 
In the framework of the $\kappa$-EED hypothesis, 
 two effects can cause 
the disagreement between the  $\kappa$ indices derived from 
the \ion{C}{2} UV lines  ($\kappa _{\rm C II}$) and those derived from [\ion{O}{3}] CELs ($\kappa _{\rm O III}$)
as they go in opposite directions.
 If the variation of the EED with position was random, 
$\kappa _{\rm C II}$ would presumably be lower than $\kappa _{\rm O III}$ since the excitation of the \ion{C}{2} UV lines 
is more sensitive to the non-thermal tail of the EED as it preferentially takes place
within the lower-$\kappa$ regions. The PN
NGC\,6153, which presents complex nebular structures, probably
belongs to this category.  
On the other hand,  deep integrated-field spectroscopy has revealed a globally increasing ADFs towards the center of PNe
\citep{G-R}. In this case, we would expect  $\kappa _{\rm C II}>\kappa _{\rm O III}$ 
since the [\ion{O}{3}] and \ion{C}{2} lines mainly trace the highly ionized inner region and the partially ionized
outer layer, respectively. This might actually be the case for the high-excitation PN NGC\,2440.

In any case, we can  draw the firm conclusion that the $\kappa$ EED cannot be present in a homogeneous manner
throughout the whole nebulae. On the other hand, current observations cannot rule out the development in PNe of
local small-scale $\kappa$ EEDs. 
From the theoretical point of view,
the main issue with the $\kappa$ EED hypothesis
is the lack so far of a plausible mechanism for pumping the non-thermal electrons in PN environments. This might be the main reason for the fact that
the $\kappa$ EED has been generally ignored in mainstream PN studies.
However, we should be aware that the
 $\kappa$ distribution is a cutting edge research area in the field of solar plasma physics although its origin still remains unclear \citep[e.g.][]{dud,dz21}. Furthermore, the other proposed solutions to the ``electron temperature and abundance
 discrepancy problem'' of PNe, such as  temperature fluctuations or  two-component models, are facing a similar
 challenge, namely, how to generate a sufficiently
  large  $t^2$ or the  hydrogen-deficient clumps? We
  therefore see no legitimate
 reason for the study of $\kappa$ distribution to be overlooked
 and undervalued by the PN research community.
 Recent studies reveal that high ADFs might be associated to the evolution of  close binary central stars \citep{wesson18} and, at
  this point, we may tentatively conjecture that a fast magnetosonic wave excited by 
 the interaction with a strong stellar magnetic fields could instantly inject energetic electrons into the 
free-electron pool, thereby forming a  $\kappa$ EED within the central regions of PNe.

We have built a toy two-EED model in an attempt to reproduce the
observations, in which, besides a regular nebular component
with a MB EED, we added a $\kappa$-EED component. In this scenario, 
the predicted  \ion{C}{2} line ratio  is given by
       \begin{equation}  \label{omg}
      r({\rm C~II})=\frac{\omega \epsilon_\kappa(2326; \kappa, T_U)+(1-\omega) \epsilon_M(2326; T_M)}{\omega \epsilon_\kappa(1335; \kappa, T_U)+(1-\omega) \epsilon_M(1335; T_M)}\,,
   \end{equation}
where  $\omega$ is the volume filling factor of the $\kappa$-EED gas, and
the emissivities of the $\kappa$- and MB-EED components,
$\epsilon_\kappa(\lambda; \kappa, T_U)$ and $\epsilon_M(\lambda; T_M)$, are derived by solving  level balance equations,
as described in Section~\ref{meth}. 
For the sake of simplicity, we have assumed that
the two components have identical density and temperature
values,  that is $T_U=T_M$ (where $T_M$ is derived from the BJ). 
Various $\kappa$ values have been found across the different solar plasmas,
for instance, 
3 in the outer solar corona, 4 in the 
outer heliosphere, 5 in the magnetotail, and 17
in the lower solar corona \citep{Livadiotisb}.
We have assumed that the $\kappa$-EED nebular component 
present values that are similar to those of solar plasmas, and  calculate 
$r$(\ion{C}{2}) for four preselected $\kappa$ values (3, 5, 10, and 20); the results are presented in Figure~\ref{Fig.6}.
It shows that with increasing  filling factor ($\omega$) of the $\kappa$-EED component, the $r$(\ion{C}{2}) versus $T_M$ relation changes
from that of a uniform MB-EED ($\omega=0$)
up to the case of a complete $\kappa$-EED ($\omega=1$).
For a given $\omega$, there is a trend that, along 
with decreasing temperature,
the $r$(\ion{C}{2}) versus $T_M$ curves get progressively closer  to
the complete $\kappa$-EED case.
This is because at such a low temperature, the MB-EED
hardly excites the \ion{C}{2} CELs unless non-thermal 
high-energy electrons exist, and thus in the low temperature case, the $r$(\ion{C}{2})
ratio from  Equ.~\ref{omg} is essentially dominated by the 
$\kappa$-EED component.

The observed PN data are  plotted in  
Figure~\ref{Fig.6}. We can see that a
very small amount of $\kappa$-EED gas with an
extremely low $\kappa$ index is sufficient to explain
the observations. For instance, we would require only
1--$5\%$ of the nebular gas to have $\kappa=3$ in order to reproduce
the observed $r$(\ion{C}{2}).
If the non-thermal gas had a $\kappa$ value of 
5 or 10, the required filling factor would need to increase up to $10\%$ or $50\%$. 
On the other hand, if one assumes $\kappa=20$, we could no longer reproduce
the lowest $r$(\ion{C}{2}) ratios. Alternatively,
 we could relax the assumption of a uniform temperature for the two components
since the ejection of non-thermal electrons into
the regular nebular gas is likely to cause $T_U$ to become $>T_M$.  
This would enhance the $\epsilon_\kappa$ value relative to $\epsilon_M$.
As a result, the inferred $\omega$ would be further reduced (see  Equ.~\ref{omg}). 
It therefore appears that the problem of the origin of $\kappa$-electrons pumping is not a very crucial issue.

\section{Conclusion}
The current work shows that the 
\ion{C}{2}$\lambda\lambda$1335,2326 UV lines can 
serve as a useful tool to investigate the EED of PNe.
The observed \ion{C}{2} 
$\lambda2326$/$\lambda$1335 intensity ratios are systematically
lower than those expected with the MB-EED hypothesis.
This  would be satisfactorily explained if we assume the $\kappa$-EED
hypothesis. We would suggest that the deviation from the MB-EED in PNe is lower than for AGNs. However, the $\kappa$-EED 
is unlikely to extend uniformly across the whole nebulae.
Using a toy model of two-EED components, we have demonstrated
that only a few percent of nebular gas in $\kappa$-EED
is sufficient to reproduce the observations.

By no means  can we state that the evidence of the existence of $\kappa$-EED in PNe is conclusive. Instead, we have shown
that the generation of non-thermal electrons can
definitely improve, rather than worsen, the agreement with the observations of \ion{C}{2} UV lines as well as other emission lines.
 The same situation occurs with the Balmer continuum discontinuity,
[\ion{O}{3}], and \ion{O}{2} observations \citep{Zhangb,Zhangc,Lin},
which supports the hypothesis that  $\kappa$-EED on a small spatial scale
is a promising solution to the ``electron temperature and abundance discrepancy problem''. 
Further observational investigations of the influence of the
$\kappa$ EED on other
spectral features of PNe  are eminently worth pursuing.

\begin{acknowledgements}
We are grateful to the anonymous referee for constructive comments that contributed to improve the manuscript. 
This work was partially based on INES data from the IUE satellite. We thank Bao-Zhi Lin for fruitful discussions.
We are grateful for financial supports from
 the National Science Foundation of China (NSFC, Grant No. 11973099) and the science research grants from the China Manned Space Project (NO. CMS-CSST-2021-A09 and CMS-CSST-2021-A10).
\end{acknowledgements}


\begin{figure*}
  \centering
  \scalebox{0.8}{\includegraphics{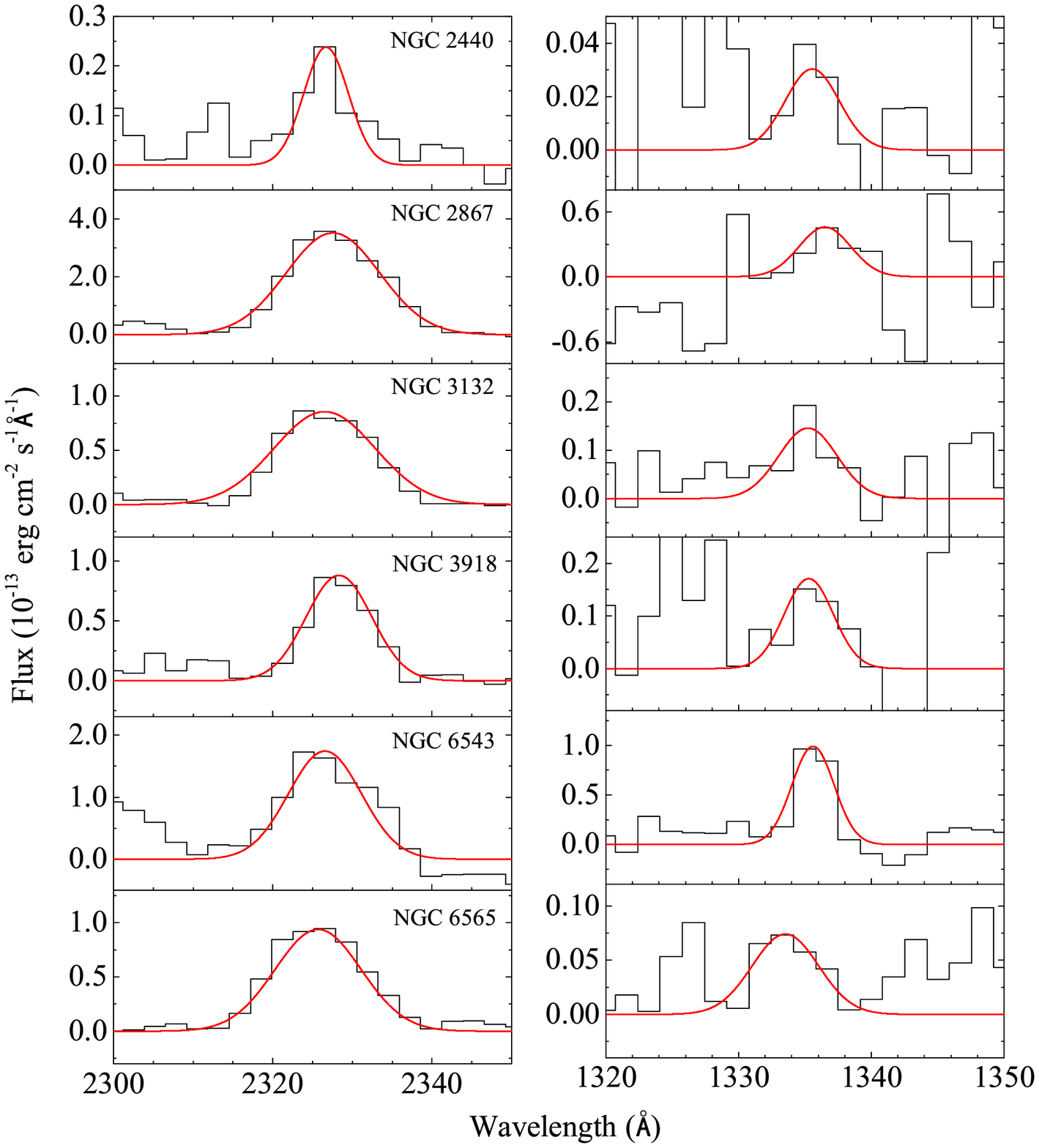}}
  \caption{The IUE spectra of the PNe, which show the \ion{C}{2}]$\lambda$2326 (left panel) and \ion{C}{2}$\lambda$1335 (right panel) features.
  The smooth curves represent the Gaussian fits.}
  \label{Fig.1}%
  \end{figure*}
    
  \begin{figure*}[htb]\addtocounter{figure}{-1}
  \centering
  \scalebox{0.8}{\includegraphics{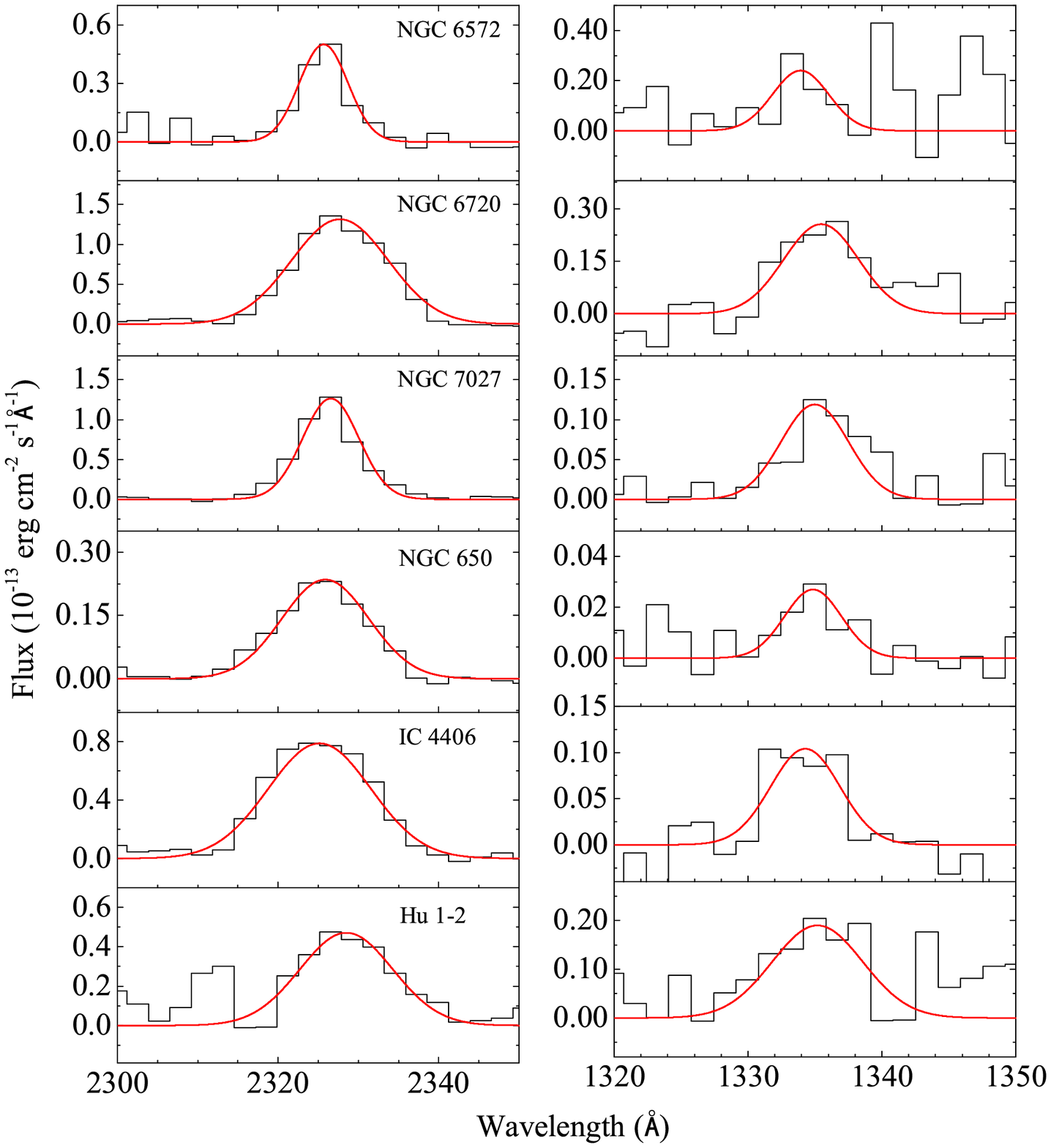}}
  \caption{-- {\emph {continued}}}
  \end{figure*}


 
 \begin{figure*}
   \centering
   \includegraphics[width=\hsize]{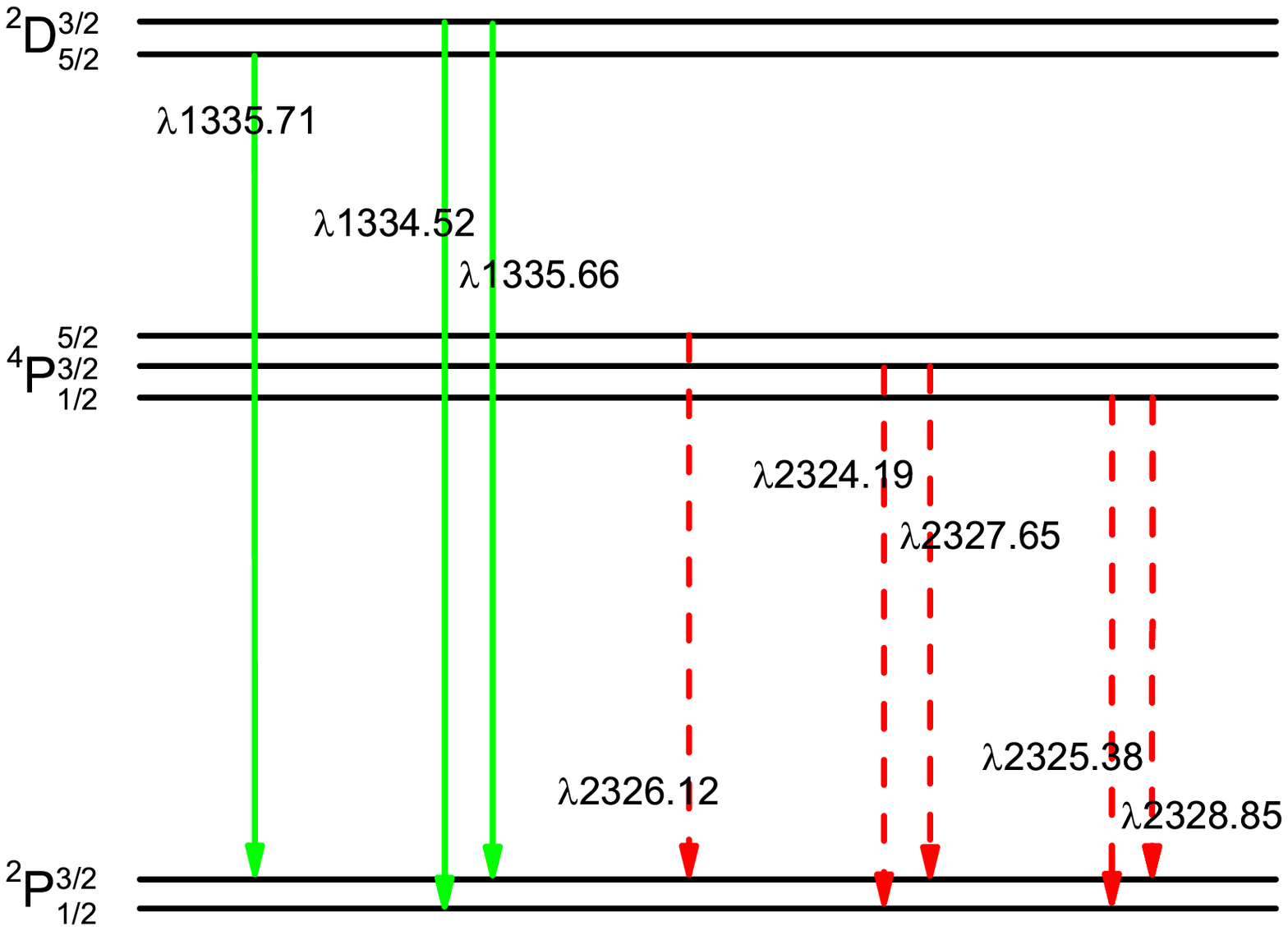}
   \caption{Configuration of the energy levels and transitions of the \ion{C}{2} ion.}
   \label{Fig.2}%
    \end{figure*}

     \begin{figure*}
   \centering
   \includegraphics[width=\hsize]{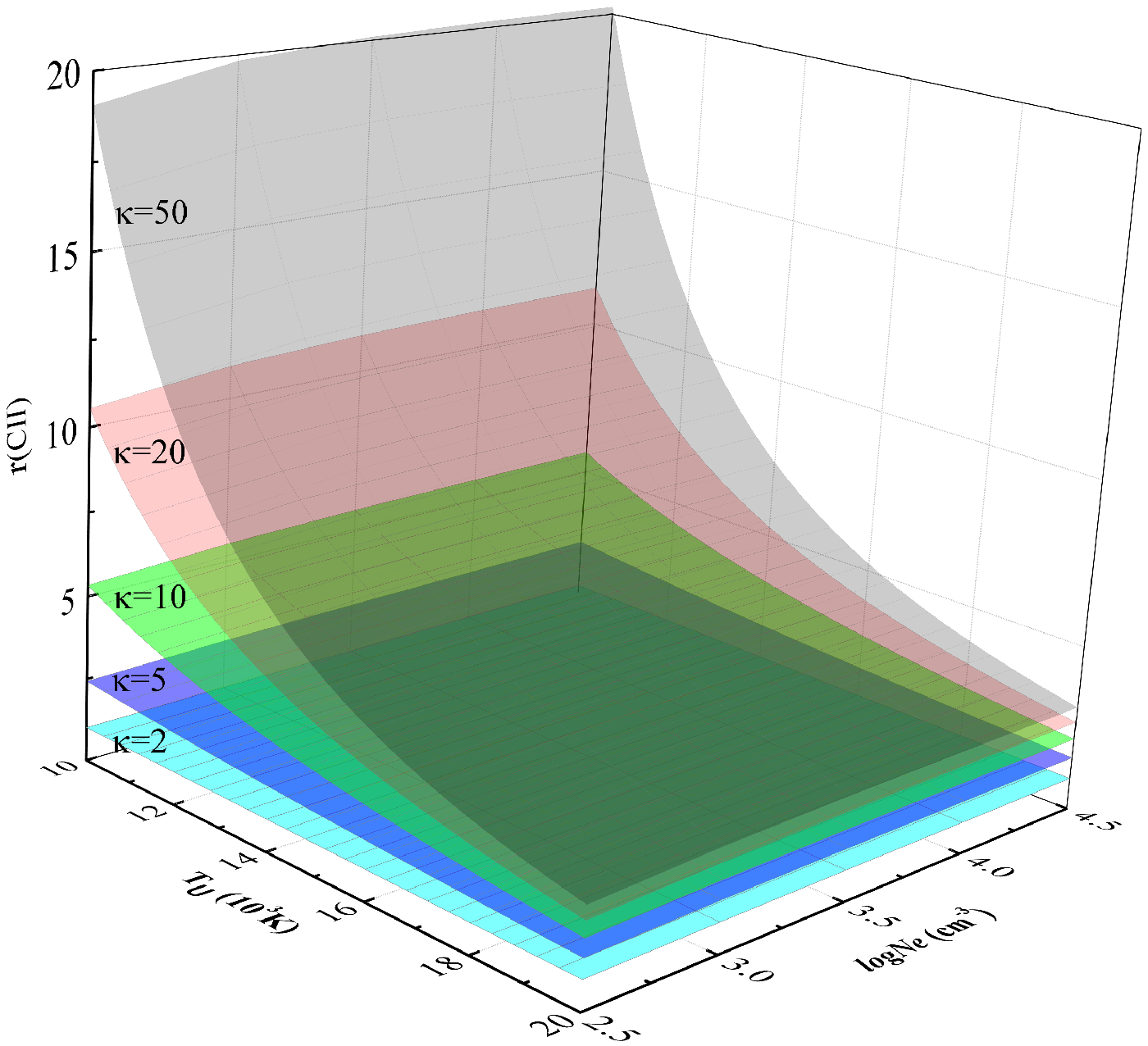}
   \caption{The theoretical \ion{C}{2} line ratio as a function of  $T_{U}$ and $N_e$ for different $\kappa$ values.}
  \label{Fig.3}%
    \end{figure*}

    \begin{figure*}
   \centering
   \includegraphics[width=\hsize]{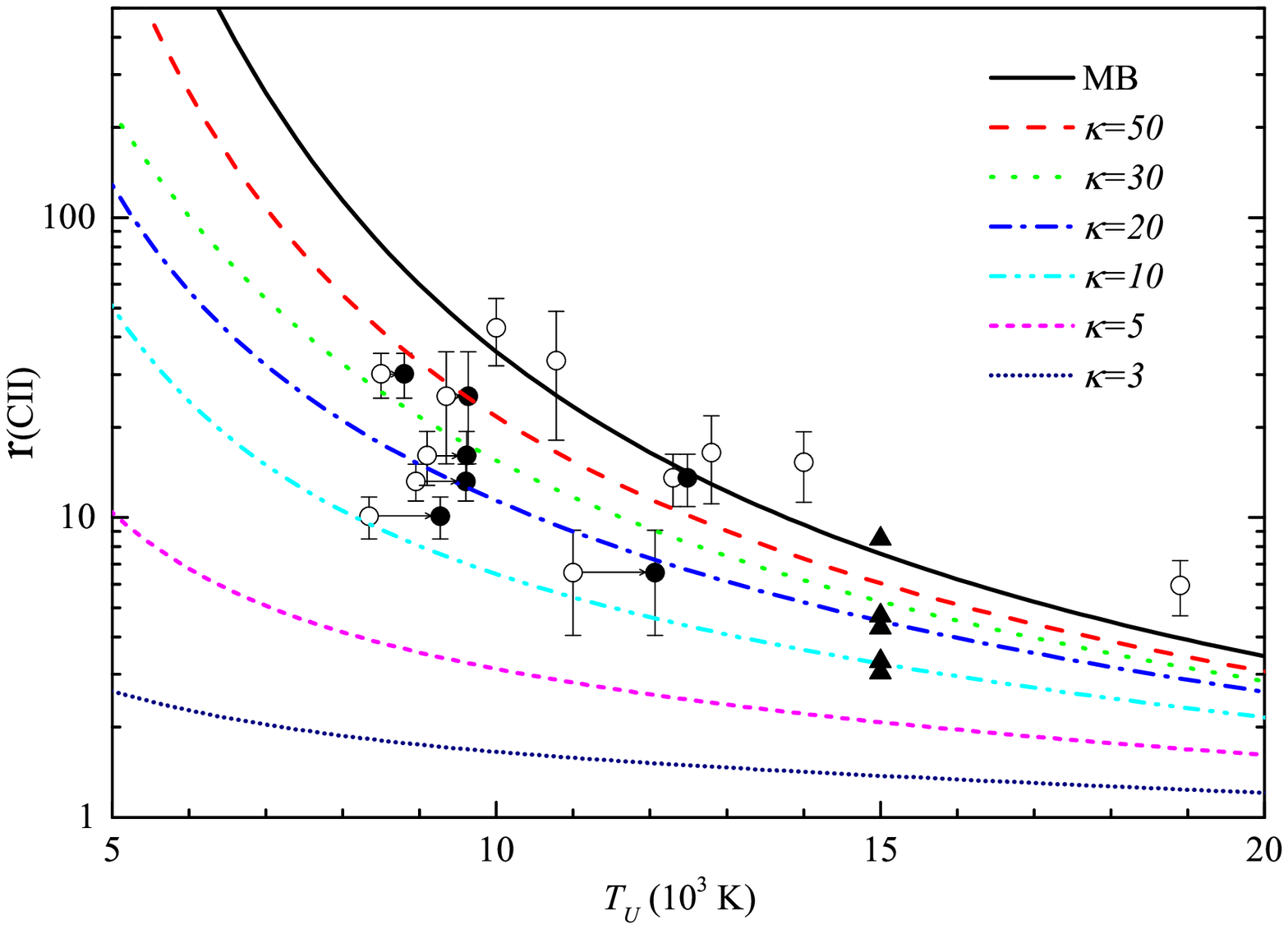}
   \caption{The 
   \ion{C}{2}]$\lambda$2326/\ion{C}{2}$\lambda$1335
   line intensity ratio vs
   $T_U$ for $N_e=1000$\,cm$^{-3}$.
   The curves indicate the theoretical predictions for
   various $\kappa$ values, among which 
   the solid line represents  the $\kappa=\infty$ case
   (i.e. the MB distribution).
   The circles with error bars are the observed values in PNe,
   where the open and filled circles represent 
   $T_{M}$(BJ) and the derived $T_U$ (see text), respectively.
    The filled triangles denote the observed
   values in quasars extracted from \citet{Vernet}, where we have assumed $T_U \equiv 1.5\times10^4$\,K.
   }
              \label{Fig.4}%
    \end{figure*}   
 
     \begin{figure*}
   \centering
   \includegraphics[width=\hsize]{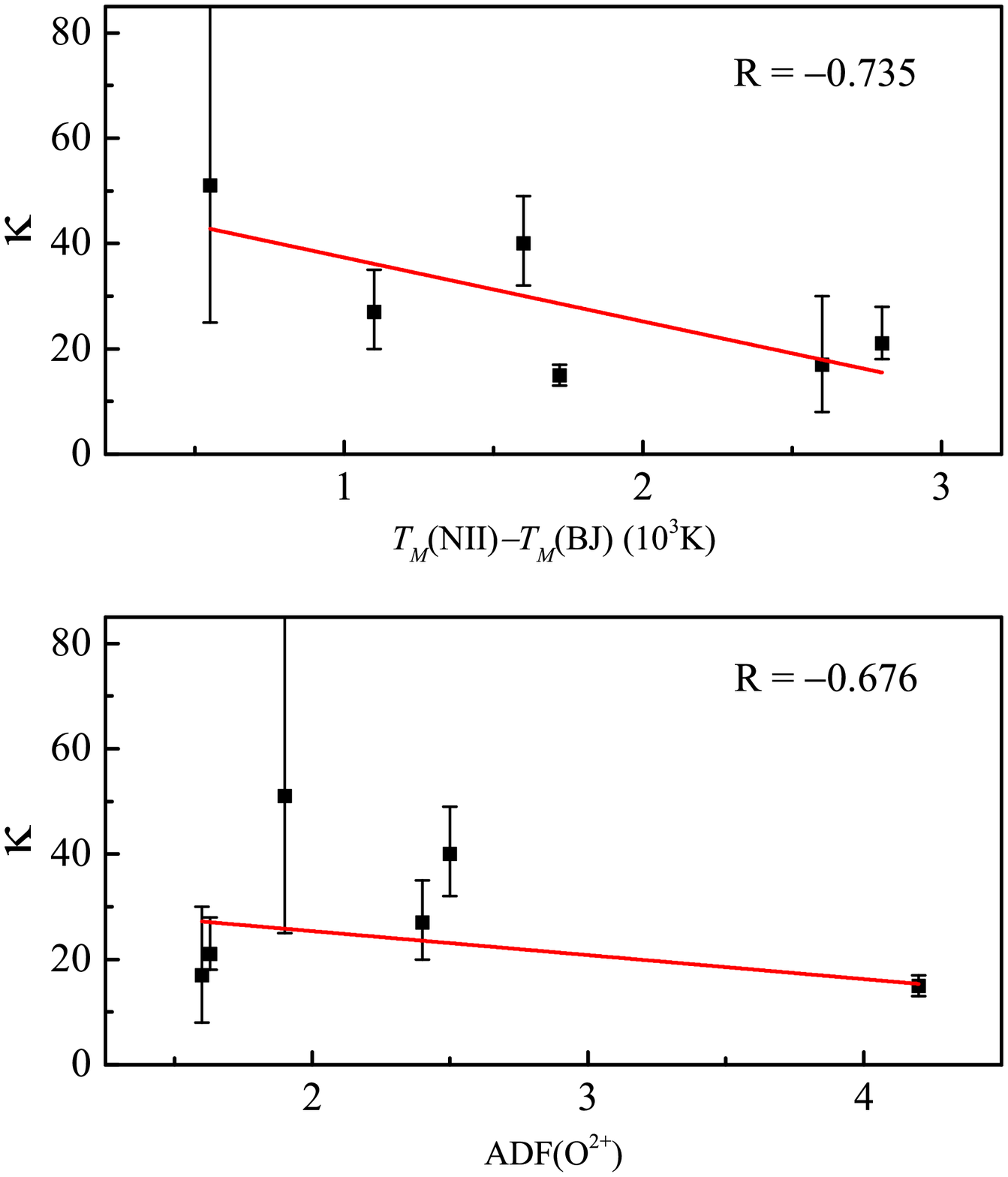}
   \caption{ The
   $\kappa$ index  vs $T_{M}$(\ion{N}{2})$-T_{M}$(BJ) (upper panel) and ADF(O$^{2+}$) (lower panel). The solid lines represent linear fittings. The correlation coefficients are shown in the up-right corners.}
              \label{Fig.5}%
    \end{figure*}

     \begin{figure*}
   \centering
   \includegraphics[width=\hsize]{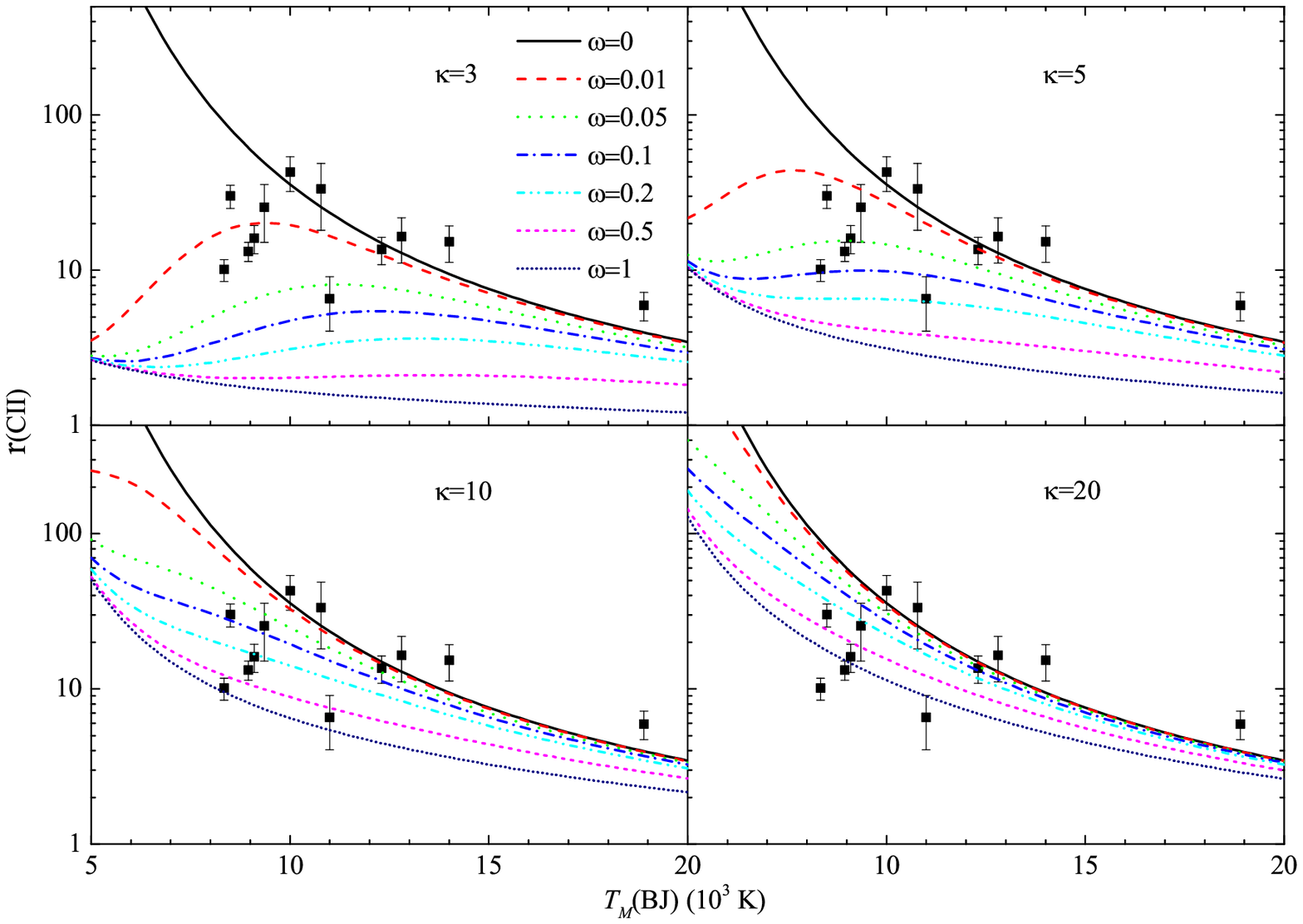}
   \caption{ The relations between $r$(\ion{C}{2}) and $T_{M}$(BJ) predicted  by the two-EED model, which 
   is composed of a regular MB-EED nebular component and a $\kappa$-EED component 
   with a $\kappa$ index of 3, 5, 10, and 20 and various filling factors ($\omega$).
  The squares with error bars represent the observed values.}
              \label{Fig.6}%
    \end{figure*}

\begin{longrotatetable}
\begin{deluxetable}{lcccccccccccc}
\tablecaption{The sample of PNe, with the flux and intensity measurements of the two \ion{C}{2} UV features,  and the diagnostic results. Note that
 the $F(\lambda)$ and $I(\lambda)$ are those of the blended features.\label{tab1}
  }
 \tabletypesize{\scriptsize}

\setlength{\tabcolsep}{2mm}

\tablehead{
    Objects & \multicolumn{2}{c}{F($\lambda$) (10$^{-13}$ $erg$ $cm^{-2}$ $s^{-1}$)}  & c(H$\beta$)$^{a}$ & \multicolumn{2}{c}{I($\lambda$) (10$^{-13}$ $erg$ $cm^{-2}$ $s^{-1}$)} & $r$(\ion{C}{2}) & $T_{M}$(\ion{N}{2}) & $T_{M}$(BJ) & ADF   & \multicolumn{2}{c}{$\kappa$}   &  $T_{U}$  \\[3pt]
        \cline{2-3}  \cline{5-6}  \cline{11-12} 
          & \ion{C}{2}] $2326$ & \ion{C}{2} $1335$ &      & \ion{C}{2}] $2326$ & \ion{C}{2} $1335$ &     &  (K)   &  (K)  &    &  This    &  ref.$^l$  &  (K)   
}
\startdata
    NGC 2440 & $1.64\pm0.25$  & $0.15\pm0.03$ & $0.42$  & 6.11 $\pm0.91$  & $0.59\pm0.13$ & $15.29\pm4.02$ & $10900$ $^{b}$ & $14000$ $^{b}$ & $5.4$ $^{d}$     &  $\infty$  &  25 & $14000$   \\[3pt]
    NGC 2867 & $51.11\pm1.44$  & $5.32\pm0.81$ & 0.43  & $196.20\pm5.51$ & $21.73\pm3.01$ & $13.22\pm1.87$ & $11750$ $^{k}$ & 8950 $^{k}$  & 1.63 $^{k}$     & \multicolumn{1}{c}{$21_{-3}^{+7}$} &  15  & 9605    \\[3pt]
    NGC 3132 & $12.28\pm0.59$  & $0.53 \pm0.24$ & 0.16  & $20.26\pm0.97$ & $0.89\pm0.41$ & $33.38\pm15.29$ & 9350 $^{c}$  & 10780 $^{c}$ & 3.5 $^{d}$      & $\infty$ &  $\infty$  &  10780   \\[3pt]
    NGC 3918 & $8.45\pm0.43$  &  $0.88 \pm0.17$  & 0.27  & $19.66\pm0.99$ & $2.12\pm0.41$ & $13.58\pm2.70$ & 10800 $^{c}$ & 12300 $^{c}$ & 1.8 $^{d}$       &  \textgreater$100$   &  \textgreater$60$ & 12300   \\[3pt]
    NGC 6543 & $25.46\pm3.66$  & $3.61 \pm0.25$  & 0.12  & $37.06\pm5.33$ & $5.35\pm0.37$ & $10.10\pm1.62$ & 10060 $^{h}$ & 8340 $^{h}$  & 4.2 $^{f}$     & \multicolumn{1}{c}{$15_{-2}^{+2}$} & $\infty$  & 9267    \\[3pt]
    NGC 6565 & $8.53\pm0.55$  & $0.39\pm0.06$  & 0.42  & $31.74\pm2.05$ & $1.54\pm0.24$ & $30.18\pm5.13$ & 10100 $^{j}$ & 8500 $^{j}$  & 1.69 $^{j}$    & \multicolumn{1}{c}{$40_{-8}^{+9}$} & 24  & 8801    \\[3pt]
    NGC 6572 & $4.02\pm0.20$  & $0.86 \pm0.32$  & 0.34  & $11.65\pm0.58$ & $2.6\pm0.99$ & $6.56\pm2.51$ & 13600 $^{e}$ & 11000 $^{e}$ & 1.6 $^{f}$      & \multicolumn{1}{c}{$17_{-9}^{+13}$} & $\infty$  & 12065   \\[3pt]
    NGC 6720 & $17.99\pm0.97$  & $1.57 \pm0.31$  & 0.29  & $44.58\pm2.40$ & $4.06\pm0.80$ & $16.10\pm3.31$ & 10200 $^{e}$ & 9100 $^{e}$  & 2.4 $^{f}$     & \multicolumn{1}{c}{$27_{-7}^{+8}$} & 32  & 9615    \\[3pt]
    NGC 7027 & $11.74\pm0.84$  & $0.87 \pm0.28$  & 1.24  & $568.08\pm40.41$ & $50.47\pm15.99$ & $16.48\pm5.35$ & 12900 $^{g}$ & 12800 $^{g}$ & 1.29 $^{g}$    & $\infty$  & \textgreater$60$ & 12800  \\[3pt]
    NGC 650 & $3.29\pm0.54$ & $0.11\pm0.02$  & 0.20  & $6.15\pm1.02$ & $0.21\pm0.04$ & $42.89\pm10.84$ & 10000:  &  10000:   &  *      & $\infty$ &  $\infty$ & 10000:    \\[3pt]
    IC 4406 & $12.63\pm3.74$  & $0.70\pm0.19$  & 0.28  & $30.34\pm8.99$ & $1.74\pm0.48$ & $25.43\pm10.32$ & 9900 $^{c}$  & 9350 $^{c}$  & 1.9 $^{d}$     & \multicolumn{1}{c}{$51_{-26}^{+50}$} & \textgreater$60$ & 9633    \\[3pt]
    Hu 1-2 & $7.03\pm1.14$  & $1.59\pm0.21$ & 0.59  & $44.52\pm7.22$ & $10.98\pm1.45$ & $5.94\pm1.24$ & 13000 $^{e}$ & 18900 $^{e}$ & 1.6 $^{f}$      & $\infty$ & $\infty$ & 18900   \\[3pt]
\enddata
   
  %
 \textbf{}
\tablenotetext{}
{Reference: $^{a}$\citet{Cahn}, $^{b}$\citet{Bernard-Salas}, $^{c, d}$\citet{Tsamis, Tsamisb},  $^{e, f}$\citet{Liuc, Liud}, $^{g}$\citet{Zhangd}, $^{h}$\citet{Wessona}, $^{i}$\citet{Wessonb}, $^{j}$\citet{Wang},
  $^{k}$\citet{Ga-Ro},
  $^{l}$\citet{Zhangc}.}

\end{deluxetable}
\end{longrotatetable}

\clearpage
\newpage
\setlength{\tabcolsep}{5mm}
\begin{longtable}{rrrrrrrrr}
\caption{The computed $r$(\ion{C}{2}).\label{A1}}\\

\hline
\hline  \\

$T_{U}$(K)  &  \multicolumn{8}{c}{$\kappa$} \\
  \cline{2-9} \\
  & 2     & 3     & 5     & 10    & 20    & 30    & 50    & $\infty$ (MB) \\ 
\hline  \\

\endfirsthead
\caption{Continued.} \\
\hline
\hline\\
$T_{U}$(K)  &  \multicolumn{8}{c}{$\kappa$} \\
  \cline{2-9} \\
   & 2     & 3     & 5     & 10    & 20    & 30    & 50    & $\infty$ (MB) \\ 
\hline  \\
\endhead
\hline
\endfoot
\hline
\endlastfoot
    5000  & 1.088  & 2.63  & 10.39  & 50.47  & 128.69  & 226.10  & 919.40  & 3685.00  \\
    5200  & 1.083  & 2.54  & 9.38  & 42.58  & 106.30  & 189.60  & 687.70  & 2583.00  \\
    5400  & 1.079  & 2.47  & 8.55  & 36.41  & 89.20  & 160.19  & 525.50  & 1858.00  \\
    5600  & 1.074  & 2.40  & 7.85  & 31.54  & 75.84  & 136.39  & 409.39  & 1367.00  \\
    5800  & 1.069  & 2.33  & 7.26  & 27.62  & 65.27  & 116.90  & 324.60  & 1027.00  \\
    6000  & 1.065  & 2.27  & 6.75  & 24.44  & 56.80  & 101.00  & 261.39  & 786.90  \\
    6200  & 1.062  & 2.22  & 6.32  & 21.82  & 49.91  & 87.88  & 213.69  & 613.29  \\
    6400  & 1.059  & 2.17  & 5.95  & 19.65  & 44.25  & 76.97  & 176.89  & 485.50  \\
    6600  & 1.057  & 2.12  & 5.63  & 17.82  & 39.53  & 67.84  & 148.19  & 389.89  \\
    6800  & 1.054  & 2.08  & 5.34  & 16.29  & 35.57  & 60.15  & 125.50  & 317.10  \\
    7000  & 1.054  & 2.04  & 5.08  & 14.96  & 32.20  & 53.63  & 107.30  & 261.00  \\
    7200  & 1.052  & 2.00  & 4.86  & 13.81  & 29.31  & 48.07  & 92.58  & 217.19  \\
    7400  & 1.052  & 1.96  & 4.65  & 12.82  & 26.81  & 43.30  & 80.50  & 182.50  \\
    7600  & 1.051  & 1.93  & 4.47  & 11.96  & 24.64  & 39.18  & 70.52  & 154.80  \\
    7800  & 1.049  & 1.90  & 4.30  & 11.18  & 22.73  & 35.60  & 62.20  & 132.39  \\
    8000  & 1.049  & 1.87  & 4.15  & 10.51  & 21.06  & 32.49  & 55.21  & 114.09  \\
    8200  & 1.047  & 1.84  & 4.01  & 9.91  & 19.57  & 29.76  & 49.28  & 99.05  \\
    8400  & 1.047  & 1.81  & 3.88  & 9.37  & 18.25  & 27.36  & 44.23  & 86.57  \\
    8600  & 1.047  & 1.79  & 3.76  & 8.88  & 17.06  & 25.23  & 39.90  & 76.14  \\
    8800  & 1.046  & 1.77  & 3.65  & 8.44  & 16.00  & 23.35  & 36.16  & 67.36  \\
    9000  & 1.044  & 1.74  & 3.55  & 8.03  & 15.03  & 21.67  & 32.91  & 59.91  \\
    9200  & 1.042  & 1.72  & 3.45  & 7.67  & 14.16  & 20.17  & 30.07  & 53.56  \\
    9400  & 1.042  & 1.70  & 3.37  & 7.33  & 13.37  & 18.82  & 27.58  & 48.11  \\
    9600  & 1.041  & 1.69  & 3.28  & 7.03  & 12.65  & 17.60  & 25.38  & 43.41  \\
    9800  & 1.039  & 1.67  & 3.20  & 6.74  & 11.99  & 16.50  & 23.44  & 39.33  \\
    10000 & 1.037  & 1.65  & 3.13  & 6.48  & 11.38  & 15.50  & 21.71  & 35.77  \\  
    10200 & 1.036  & 1.63  & 3.06  & 6.24  & 10.83  & 14.60  & 20.17  & 32.66  \\
    10400 & 1.034  & 1.62  & 2.99  & 6.01  & 10.31  & 13.77  & 18.79  & 29.92  \\
    10600 & 1.031  & 1.60  & 2.93  & 5.80  & 9.84  & 13.02  & 17.55  & 27.51  \\
    10800 & 1.029  & 1.59  & 2.87  & 5.61  & 9.40  & 12.33  & 16.43  & 25.36  \\
    11000 & 1.028  & 1.58  & 2.82  & 5.42  & 8.99  & 11.69  & 15.41  & 23.45  \\
    11200 & 1.026  & 1.56  & 2.76  & 5.25  & 8.61  & 11.11  & 14.49  & 21.75  \\
    11400 & 1.024  & 1.55  & 2.71  & 5.09  & 8.25  & 10.57  & 13.66  & 20.22  \\
    11600 & 1.021  & 1.54  & 2.66  & 4.94  & 7.92  & 10.07  & 12.89  & 18.85  \\
    11800 & 1.019  & 1.53  & 2.62  & 4.79  & 7.61  & 9.61  & 12.19  & 17.61  \\
    12000 & 1.016  & 1.51  & 2.57  & 4.66  & 7.32  & 9.18  & 11.55  & 16.49  \\
    12200 & 1.014  & 1.50  & 2.53  & 4.53  & 7.05  & 8.78  & 10.96  & 15.47  \\
    12400 & 1.011  & 1.49  & 2.49  & 4.41  & 6.80  & 8.41  & 10.42  & 14.55  \\
    12600 & 1.008  & 1.48  & 2.45  & 4.29  & 6.56  & 8.07  & 9.92  & 13.71  \\
    12800 & 1.006  & 1.47  & 2.41  & 4.18  & 6.33  & 7.74  & 9.46  & 12.94  \\
    13000 & 1.003  & 1.46  & 2.38  & 4.08  & 6.12  & 7.44  & 9.03  & 12.23  \\
    13200 & 1.001  & 1.45  & 2.34  & 3.98  & 5.92  & 7.16  & 8.63  & 11.58  \\
    13400 & 0.999  & 1.44  & 2.31  & 3.88  & 5.73  & 6.89  & 8.26  & 10.99  \\
    13600 & 0.997  & 1.43  & 2.28  & 3.79  & 5.55  & 6.64  & 7.91  & 10.44  \\
    13800 & 0.994  & 1.42  & 2.25  & 3.70  & 5.38  & 6.41  & 7.59  & 9.93  \\
    14000 & 0.992  & 1.41  & 2.22  & 3.62  & 5.22  & 6.18  & 7.29  & 9.46  \\
    14200 & 0.989  & 1.41  & 2.19  & 3.54  & 5.06  & 5.97  & 7.01  & 9.03  \\
    14400 & 0.987  & 1.40  & 2.16  & 3.47  & 4.92  & 5.78  & 6.74  & 8.63  \\
    14600 & 0.984  & 1.39  & 2.13  & 3.39  & 4.78  & 5.59  & 6.50  & 8.25  \\
    14800 & 0.982  & 1.38  & 2.11  & 3.32  & 4.64  & 5.41  & 6.26  & 7.90  \\
    15000 & 0.980  & 1.37  & 2.08  & 3.26  & 4.52  & 5.25  & 6.04  & 7.57  \\
    15200 & 0.977  & 1.37  & 2.06  & 3.19  & 4.40  & 5.09  & 5.84  & 7.27  \\
    15400 & 0.975  & 1.36  & 2.03  & 3.13  & 4.29  & 4.94  & 5.64  & 6.98  \\
    15600 & 0.973  & 1.35  & 2.01  & 3.07  & 4.18  & 4.80  & 5.46  & 6.72  \\
    15800 & 0.970  & 1.34  & 1.99  & 3.01  & 4.07  & 4.66  & 5.28  & 6.47  \\
    16000 & 0.968  & 1.33  & 1.96  & 2.96  & 3.97  & 4.53  & 5.12  & 6.23  \\
    16200 & 0.966  & 1.33  & 1.94  & 2.90  & 3.88  & 4.41  & 4.96  & 6.01  \\
    16400 & 0.964  & 1.32  & 1.92  & 2.85  & 3.79  & 4.29  & 4.82  & 5.80  \\
    16600 & 0.962  & 1.31  & 1.90  & 2.80  & 3.70  & 4.18  & 4.68  & 5.60  \\
    16800 & 0.960  & 1.30  & 1.88  & 2.75  & 3.61  & 4.07  & 4.54  & 5.41  \\
    17000 & 0.957  & 1.30  & 1.86  & 2.71  & 3.53  & 3.97  & 4.41  & 5.24  \\
    17200 & 0.955  & 1.29  & 1.84  & 2.66  & 3.45  & 3.87  & 4.29  & 5.07  \\
    17400 & 0.953  & 1.28  & 1.82  & 2.62  & 3.38  & 3.78  & 4.18  & 4.92  \\
    17600 & 0.951  & 1.28  & 1.80  & 2.58  & 3.31  & 3.69  & 4.07  & 4.77  \\
    17800 & 0.949  & 1.27  & 1.79  & 2.54  & 3.24  & 3.60  & 3.96  & 4.63  \\
    18000 & 0.947  & 1.26  & 1.77  & 2.50  & 3.17  & 3.52  & 3.86  & 4.49  \\
    18200 & 0.946  & 1.26  & 1.75  & 2.46  & 3.11  & 3.44  & 3.77  & 4.36  \\
    18400 & 0.944  & 1.25  & 1.74  & 2.42  & 3.05  & 3.36  & 3.67  & 4.24  \\
    18600 & 0.942  & 1.25  & 1.72  & 2.38  & 2.99  & 3.29  & 3.59  & 4.13  \\
    18800 & 0.940  & 1.24  & 1.70  & 2.35  & 2.93  & 3.22  & 3.50  & 4.01  \\
    19000 & 0.938  & 1.23  & 1.69  & 2.31  & 2.88  & 3.15  & 3.42  & 3.91  \\
    19200 & 0.936  & 1.23  & 1.67  & 2.28  & 2.82  & 3.09  & 3.35  & 3.81  \\
    19400 & 0.935  & 1.22  & 1.66  & 2.25  & 2.77  & 3.03  & 3.27  & 3.71  \\
    19600 & 0.933  & 1.22  & 1.64  & 2.22  & 2.72  & 2.97  & 3.20  & 3.62  \\
    19800 & 0.931  & 1.21  & 1.63  & 2.19  & 2.67  & 2.91  & 3.13  & 3.53  \\
    20000 & 0.929  & 1.20  & 1.62  & 2.16  & 2.63  & 2.85  & 3.07  & 3.45  \\

\end{longtable}

\end{document}